\documentclass[journal]{vgtc}

\onlineid{1461}
\vgtccategory{Research}
\vgtcpapertype{Technique}

\title{HeedVision: Attention Awareness in\\ Collaborative Immersive Analytics Environments}

\author{
    \authororcid{Arvind Srinivasan}{0000-0002-3409-6077},
    \authororcid{Niklas Elmqvist}{0000-0001-5805-5301}
}

\authorfooter{
\item Arvind Srinivasan, and Niklas Elmqvist are with Aarhus University in Aarhus, Denmark. E-mail: \{arvind, elm\}@cs.au.dk. 
}

\shortauthortitle{Attention-Aware Visualization}

\abstract{%
  \textit{Group awareness}---the ability to perceive the activities of collaborators in a shared space---is a vital mechanism to support effective coordination and joint data analysis in collaborative visualization.
  We introduce \textit{collaborative attention-aware visualizations} (CAAVs) that track, record, and revisualize the collective attention of multiple users over time.
  We implement this concept in \textsc{HeedVision}, a standards-compliant WebXR system built with React Three Fiber that runs on modern AR/VR headsets, and complement it with proof-of-concept implementations covering the remaining three quadrants of our design space---varying presentation (embedded vs.\ separated) and situatedness (world space vs.\ camera space).
  Through a mixed-methods exploratory study where pairs of co-located analysts performed visual search tasks in a shared immersive AR environment, we investigate how attention revisualization affects collaborative coordination in immersive analytics.
  Our results show that CAAVs improve spatial coordination, search efficiency, and task load distribution among collaborators, though benefits vary by context, favoring abstract environments lacking natural landmarks. 
  This work extends attention awareness to multi-user settings and provides empirical evidence for its context-dependent benefits in collaborative immersive analytics environments.
}


\teaser{
    \centering
    \resizebox{\linewidth}{!}{\includegraphics[alt={HeedVision Teaser},width=\linewidth]{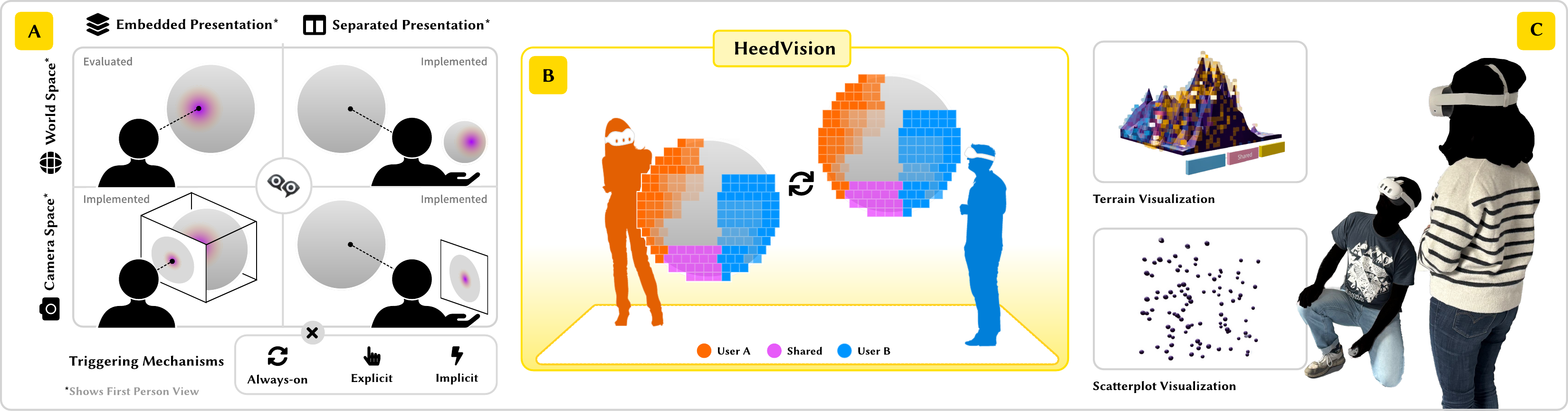}}
    \caption{\textbf{HeedVision overview.}
    We propose a (A) Collaborative Attention-Aware Visualization design space covering attention capture, presentation, situatedness, and triggering.
    (B) HeedVision implements this in a co-located mixed reality setting using voxel-based coverage maps for shared attention.
    (C) We evaluate it through a treasure hunt task across terrain and scatterplot visualizations with and without attention cues, revealing context-dependent benefits in abstract environments lacking natural reference points.}
    \label{fig:teaser}
}

\graphicspath{{figures/}{pictures/}{images/}{./}}

\usepackage[svgnames, dvipsnames]{xcolor}

\usepackage{enumitem}

\usepackage{svg}

\usepackage{fontawesome5}

\usepackage{booktabs}

\usepackage{listings}

\usepackage{colortbl}

\usepackage{tikz}

\usepackage{soul}
\definecolor{new-green}{rgb}{0.104,0.667,0.229}
\newcommand{\del}[1]{} 

\definecolor{insightbg}{RGB}{250,250,250}
\definecolor{insightborder}{RGB}{200,200,200}
\definecolor{insighttext}{RGB}{70,70,70}

\newcommand{\researchgap}[1]{%
    \tikz[baseline=(key.base)]{%
    \node[
      draw,
      rounded corners,
      inner sep=2pt,
      outer sep=0pt,
      text=insighttext,
      fill=insightbg,
      draw=insightborder,
      font=\footnotesize\scshape,
      name=key
    ] {Research Gap};
  }%
  \space\textit{#1}
}

\makeatletter
\newlength{\mylength}

\newsavebox{\mybox}

\makeatother

\begin{document}

\maketitle


\section{Introduction}

Attention is a scarce commodity.
Human cognition uses attention as a selection mechanism that determines which sensory inputs are prioritized for processing at any given moment~\cite{Treisman1980}.
In visualization, attention helps the analyst focus on areas of interest across charts to address the current analysis goal~\cite{DBLP:journals/tvcg/SrinivasanEBRE25}.
This is particularly important in collaborative visualization~\cite{Isenberg2011}, as it enables multiple analysts to jointly make sense of complex data. However, effective collaboration demands that each participant maintain awareness of what their partners are doing.
This \textit{group awareness} (the ability to perceive the activities of collaborators in a shared space~\cite{Gutwin1998a,DBLP:conf/cscw/GutwinG98}) becomes especially challenging in immersive 3D environments, where multiple viewpoints, distributed spatial attention, and real-time coordination create unique obstacles.
Prior work has shown how AR/VR environments change the dynamics of collaborative exploration and sensemaking~\cite{Marriott2018, Ens2021}, and how gaze, gesture, and embodied cues can support group awareness~\cite{Saffo_Eyes_Shoes,DBLP:series/lncs/BillinghurstCBM18}.
More recent efforts have explored representing collaborators' focus of attention on desktop and large-display systems~\cite{Jianu2025,10.1145/3715669.3725871}.
Yet few systems directly capture and represent collaborators' attention over time in immersive environments, leaving a gap at the intersection of attention-aware visualization and immersive collaboration.

Attention-aware visualizations (AAVs)~\cite{DBLP:journals/tvcg/SrinivasanEBRE25} address part of this challenge for individual users: they measure a single viewer's attention over time and modify the display accordingly to encourage exploration or support reflection.
In this paper, we extend AAVs to multi-analyst settings in what we call \textit{Collaborative Attention-Aware Visualizations} (CAAVs).
While this extension builds on the same three-component structure (capture, record, display), it introduces fundamentally new design challenges with no single-user analog.
For capture, we must distinguish between users while tracking attention simultaneously.
For recording, we maintain identity-aware attention accumulators and introduce aggregation methods (sum, maximum, difference, count) to represent collective patterns.
For display, we must encode both attention intensity and user identity while supporting coordination rather than merely self-reflection.
These challenges motivate a systematic design space for CAAVs addressing three research questions: how to \textbf{capture} collective attention in real time (RQ1), how to \textbf{record} it over time (RQ2), and how to \textbf{display} it effectively (RQ3).

To probe this design space through a concrete instantiation, we developed \textsc{HeedVision}, an immersive analytics environment that tracks and revisualizes the attention of collaborating pairs.
\textsc{HeedVision} embodies the world-space embedded configuration of our design space: head-based attention capture for accessibility across consumer devices, voxel-based recording with temporal decay, and explicit world-embedded display with color-per-user encoding.
Built using React Three Fiber with WebXR support, the system runs on consumer-level AR/VR headsets such as the Meta Quest 3.
We complement \textsc{HeedVision} with proof-of-concept implementations of the remaining three quadrants (world-space separated, camera-space embedded, and camera-space separated), illustrating the breadth of the design space.
We evaluated \textsc{HeedVision} through a mixed-methods exploratory study with eight co-located pairs (16 participants) performing collaborative visual search tasks in a shared augmented reality environment.
To understand whether benefits depend on visualization context, we tested CAAVs across both abstract discrete environments (3D scatterplots) and continuous environments with natural landmarks (terrain visualizations).
Our findings suggest that CAAV effectiveness depends on visualization context: attention visualization appeared to improve spatial coordination and reduce redundant exploration in abstract environments lacking natural reference points, but introduced potential interference in environments where existing landmarks already supported coordination.

We claim the following contributions:
\begin{itemize}
    \item[(1)] A design space for collaborative attention-aware visualization, extending AAV to multi-user immersive settings through three research questions that organize capture, recording, and display alternatives for collective attention;
    \item[(2)] \textsc{HeedVision}, a WebXR research prototype built with React Three Fiber that instantiates the world-space embedded configuration, complemented by proof-of-concept implementations of the remaining three design space quadrants that illustrate the framework's breadth, though these use local synchronization rather than networked collaboration and have not been formally evaluated; and
    \item[(3)] Empirical evidence from an exploratory study showing when collective attention visualization helps versus hinders collaborative coordination, revealing context-dependent benefits that depend on the structure of the visualization environment.
\end{itemize}

All design decisions, implementation parameters, and evaluation materials are documented to enable replication and extension.
Supplementary materials are available at \href{https://osf.io/frsp8/?view_only=0e5daa9f2e0246aaadb4601fbdf24ca1}{\texttt{osf.io/frsp8}} (anonymized).

\section{Background}

Our work extends attention-aware visualization to collaborative immersive settings.
Below we review the foundations on which we build.

\subsection{Group Awareness in Collaborative Visualization}

Visual attention determines which elements of a complex visualization are prioritized for cognitive processing~\cite{Treisman1980}.
Visualization designers leverage this through pre-attentive features~\cite{Healey1999} and visual saliency models~\cite{DBLP:journals/cgf/JanickeC10, matzen18visualsaliency} that predict and guide where viewers direct their focus.
Eye tracking has provided empirical insights into these attentional patterns across different visualization types~\cite{Burch_eyeTracking_11, Kim_eyeTracker_12, DBLP:conf/etra/BlascheckJKBE16, Kurzhals_16, Burch2021}.

While individual attention is well studied, collaborative visualization introduces the additional challenge of \textit{group awareness}: understanding others' activities within a shared workspace~\cite{Gutwin1998a, DBLP:conf/cscw/GutwinG98}.
Group awareness is essential for establishing \textit{common ground}: the mutual knowledge, beliefs, and assumptions that underpin effective coordination~\cite{Clark1991}.
The benefits of such collaboration are well documented: Mark et al.~\cite{Mark2005} and Balakrishnan et al.~\cite{Balakrishnan2008} both found significant improvements when analysts worked together using shared visual representations.

Several techniques support group awareness in collaborative settings: radar overviews~\cite{Gutwin1996a}, embodied cues like collaborators' arms~\cite{Tang2006b} and fingertips~\cite{Hugin}, and social navigation techniques~\cite{Dourish1994} that record and visualize user presence and activity~\cite{10.1145/3715669.3725871}.
Scented Widgets~\cite{Willett2007} embed usage traces directly within interface components, while Matejka et al.'s Patina~\cite{Matejka2013Patina} overlays dynamic heatmaps of usage patterns on controls; both demonstrate how aggregated interaction data can convey collective behavior.
Collaborative brushing techniques~\cite{Isenberg2009, Hajizadeh2013} reveal others' selections across datasets, and Saffo et al.'s ``Eyes and Shoes''~\cite{Saffo_Eyes_Shoes} technique provides asymmetric awareness cues between VR and desktop.

A related dimension of awareness is \textit{collective data coverage}: ensuring the group examines all relevant data without duplication.
Sarvghad and Tory demonstrated that visualizing dimension coverage increases exploration breadth~\cite{Sarvghad2016} and reduces work duplication in asynchronous collaboration~\cite{Sarvghad2015}.
Badam et al.~\cite{DBLP:conf/graphicsinterface/BadamZWEE17} extended this with a ``team-first'' approach that balances individual with collective sensemaking.
This notion of coverage tracking motivates our recording dimension (Section~3).

\researchgap{Most collaborative systems focus on discrete interaction events rather than continuous attention, and few leverage attention in immersive 3D where spatial coordination is paramount.}

\subsection{Gaze and Attention Tracking}

Addressing continuous attention in immersive environments requires reliable head tracking.
Gaze serves both as an implicit indicator of visual attention and as an explicit input mechanism.
Since Bolt's seminal gaze-orchestrated windows~\cite{DBLP:conf/siggraph/Bolt81}, researchers have explored gaze for target acquisition~\cite{DBLP:conf/chi/StellmachD12}, multimodal interaction~\cite{DBLP:conf/ngca/StellmachSND11, DBLP:conf/uist/PfeufferACG14, DBLP:conf/sui/PfeufferMMG17}, and cross-device interaction~\cite{DBLP:conf/interact/TurnerABSG13}.
A persistent challenge is the ``Midas touch'' problem~\cite{DBLP:conf/chi/Jacob90}, where every glance triggers an action; multimodal approaches that require secondary confirmation~\cite{DBLP:conf/ngca/StellmachSND11} and explicit triggering mechanisms can mitigate this.
Gaze control is now central to consumer XR devices such as the Apple Vision Pro and Android XR.

In immersive settings where eye trackers are unavailable or impractical, head orientation serves as a proxy for visual attention.
Studies of viewing behavior in 360\textdegree{} video and VR have shown that head and gaze directions are strongly correlated: Sitzmann et al.~\cite{Sitzmann2018} found that the mean gaze direction relative to head orientation is 13.85\textdegree{} $\pm$ 11.73\textdegree{}, indicating strong coupling between head and gaze movements, and David et al.~\cite{David2018} confirmed this coupling in active VR exploration.
This correlation is sufficiently strong for tasks where conveying general areas of attention matters more than precise fixation points; as is the case for collaborative spatial coordination, knowing approximately where a partner is looking enables effective division of labor.

\researchgap{While gaze and head tracking are well established for single users, using these signals to support awareness across multiple collaborators in immersive environments is largely unexplored.}

\subsection{Attention-Aware Visualization}

Attention-aware visualizations (AAVs) represent a class of systems that track, analyze, and respond to user attention patterns.
Srinivasan et al.~\cite{DBLP:journals/tvcg/SrinivasanEBRE25} formalized this concept with a framework organized around three components: how attention is captured, how it is recorded over time, and how the visualization adapts in response.
They distinguish between \textit{data-agnostic} approaches that track attention as spatial coordinates without knowledge of the underlying data, and \textit{data-aware} approaches that associate attention with marks or semantic components.

Building on this foundation, Jianu et al.~\cite{Jianu2025} assembled a comprehensive design framework and research agenda for gaze-aware visualizations, distilling research challenges and providing practical guidance for integrating eye-tracking into visualization systems.
Earlier systems demonstrated these principles in practice: Exploration Awareness~\cite{Exploration_Awareness_09} tracked user exploration paths through information spaces, while adaptive annotations~\cite{Kim_eyeTracker_12} adjusted to viewing patterns.
In immersive environments, Lu et al.~\cite{lu2014subtleCues} investigated subtle visual cues for directing attention in augmented reality, finding that appropriate cue designs can guide visual search without disrupting the primary task.

Beyond tracking and recording, display techniques can communicate attention history without persistent clutter.
Baudisch et al.~\cite{Baudisch2006Phosphor} introduced phosphor transitions, afterglow effects that fade over time to convey interface changes; this ephemeral approach offers relevant design alternatives for attention-aware revisualization.

However, existing AAV systems share a fundamental limitation: they are designed for a single user.
They track one viewer's attention and adapt the display for that viewer alone.
In collaborative settings, the challenge shifts from individual reflection to collective coordination; analysts need to know not only where \textit{they} have looked, but where their partners have been, where coverage gaps remain, and how to divide a shared analytical space.
These collaborative requirements motivate our extension of AAVs to multi-user settings: Collaborative Attention-Aware Visualizations (CAAVs).

\researchgap{The gap between well-developed single-user AAV approaches and the requirements of collaborative IA motivates extending attention awareness to multi-user settings, where coordination, coverage, and shared understanding become central design goals.}


\begin{figure}[htb]
    \centering
    \includegraphics[alt={A design space framework organized as a 2-by-2 grid with four quadrants showing different approaches to collaborative attention-aware visualization. The vertical axis distinguishes between World Space (top row) and Camera Space (bottom row), while the horizontal axis separates Embedded Presentation (left column) from Separated Presentation (right column). Each quadrant depicts a user silhouette viewing attention visualizations, represented as gray spheres with orange glow indicating focus points connected by dashed lines to the user's gaze direction. The top-left quadrant (World Space, Embedded Presentation) is highlighted in yellow and marked as 'Implemented', showing attention cues integrated directly into the 3D environment. The top-right quadrant shows attention information in a separate floating display. The bottom-left quadrant displays attention within a 3D perspective view box in camera space. The bottom-right quadrant shows a separated 2D panel with attention information. A central icon with eye symbols connects all quadrants. Below the grid, three triggering mechanisms are shown with icons: Always-on (circular arrows), Explicit (pointing hand), and Implicit (lightning bolt). A note indicates all views show the first-person perspective.},width=0.9\linewidth]{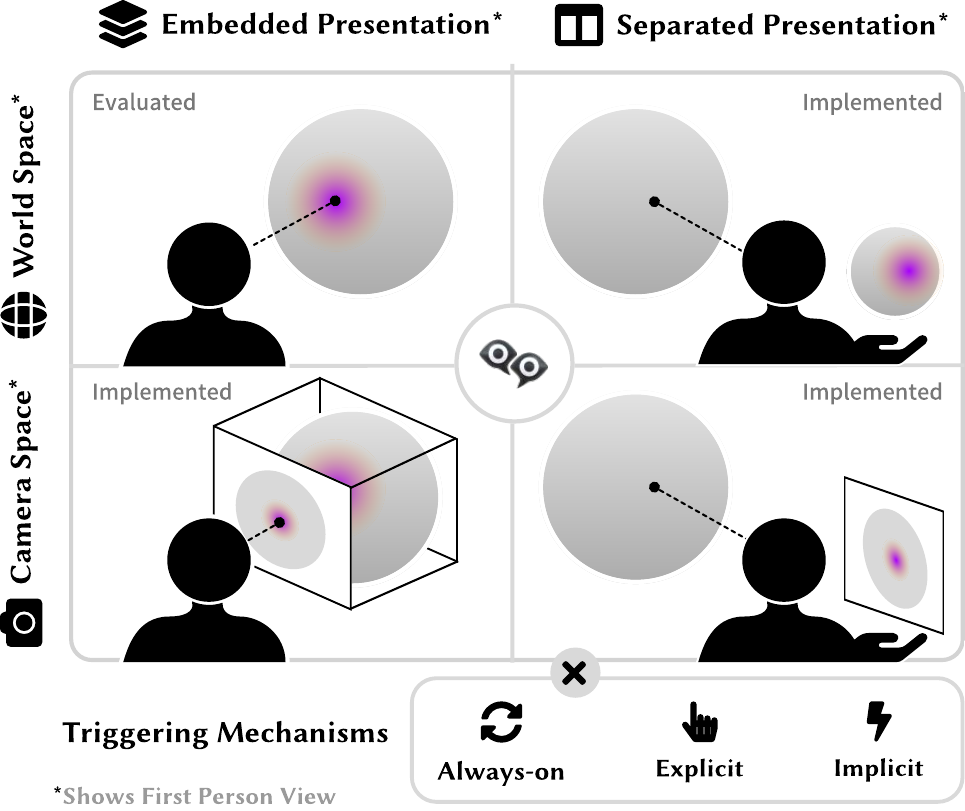}
    \caption{\textbf{Collaborative attention-aware visualization framework.}
    This framework organizes design choices across three key dimensions:
    (1) \textit{presentation} distinguishes between embedding attention directly into the visualization versus separating it into distinct views or overlays;
    (2) \textit{situatedness} determines whether attention is rendered in world space (anchored to the environment) or camera space (following the user's viewport); and
    (3) \textit{triggering} specifies when attention visualization appears: continuously (always-on), upon user request (explicit), or based on system-detected context (implicit).
    The quadrant formed by presentation and situatedness (top) illustrates the four primary design configurations, while triggering mechanisms (bottom) can be applied to any.}
    \label{fig:design-space}
\end{figure}

\section{Collaborative Attention-Aware Visualization}

Srinivasan et al.~\cite{DBLP:journals/tvcg/SrinivasanEBRE25} defined attention-aware visualizations (AAVs) as displays that track a single viewer's attention and adapt accordingly.
We extend this concept to collaborative settings: \textit{Collaborative Attention-Aware Visualizations} (CAAVs) track, record, and revisualize the attention of multiple users engaged in joint analysis tasks (Figure~\ref{fig:design-space}).
This extension preserves AAV's three-component structure but introduces collaborative dimensions at each level:

\begin{itemize}
    \item[RQ1] How should the attention of multiple viewers on a visualization be \textbf{captured in real time}?
    \item[RQ2] How should the attention of multiple viewers on a visualization be \textbf{recorded over time}?
    \item[RQ3] How should the attention of multiple viewers on a visualization be \textbf{displayed}?
\end{itemize}

We present this design space not merely as scaffolding for one system, but as a generative framework that enables systematic exploration of collaborative attention visualization.
Just as Brehmer and Munzner's~\cite{DBLP:journals/tvcg/BrehmerM13} task typology organizes the space of visualization tasks to guide design, our framework organizes the collaborative attention design space to help researchers identify unexplored configurations and reason about tradeoffs.
\textsc{HeedVision} (Section~\ref{sec:impl}) instantiates the world-space embedded quadrant and serves as the focus of our evaluation; we additionally provide implementations of the remaining three quadrants to demonstrate the framework's generative capacity.

\subsection{Measuring Collective Attention}

If attention represents a person's focus on specific elements, then \textit{collective attention} extends this across multiple analysts.
Involving multiple agents yields concepts with no single-user counterpart:

\begin{itemize}
    \item[\faUsers] \textit{Joint Attention}, when multiple users focus on one element;
    \item[\faProjectDiagram] \textit{Distributed Attention}, when analysts distribute their focus; 
    \item[\faHistory] \textit{Sequential Attention}, when analysts build on each other's focus; 
    \item[\faSearchPlus] \textit{Cumulative Attention}, as the collective span of collaborators' attention. 
\end{itemize}

Since attention cannot be directly measured, we must use observable behaviors as proxies.
In collaborative settings, these proxies must work for multiple users simultaneously: tapping or clicking; pointer movement; head direction as an indication of gaze; and eye movement for precise focus.
Each offers different tradeoffs between precision, implementation complexity, and intrusiveness.
Measuring across multiple users also introduces challenges of distinguishing between users, aligning interaction events to a common timeline, managing scalability as team size increases, and accommodating users on different devices.

\subsection{Recording Collective Attention}

CAAVs can take two fundamentally different approaches to recording attention: \textit{data-agnostic} (no knowledge of the underlying visualization) and \textit{data-aware} (specific knowledge of visualization geometry).

\paragraph{Data-agnostic.}
This approach tracks attention as coordinates or regions without knowledge of the underlying visualization elements; in other words, the system does not know what the visualization represents, only that the user's attention is moving over it.
Data-agnostic approaches are more general but limit attention interpretation. 

\paragraph{Data-aware.}
This method associates attention with specific visualization marks, data points, or semantic components.
Because of this knowledge, data-aware approaches enable richer analysis but require deeper integration with the visualization system.

\paragraph{Representing collective attention.}
Where AAV maintains a single attention accumulator per target, CAAVs must maintain separate accumulators for each user on each target.
These accumulators start at zero, increase as a user directs attention to the target, and can be combined to represent collective attention using various aggregation methods: \textit{sum} (total collective attention); \textit{maximum} (highest individual attention); \textit{difference} (balance of attention between members); and \textit{count} (number of analysts who attended to the target).
The choice of aggregation affects what collaborative patterns become visible; sum emphasizes total coverage, while difference highlights division of labor.

\paragraph{Collective attention decay.}
Human memory decays naturally over time, and recorded attention values should gradually decrease to reflect fading relevance.
Decay is particularly important in collaborative settings where attention patterns evolve as participants respond to each other's discoveries and insights.

\subsection{Revisualizing Collective Attention}

As Srinivasan et al.~\cite{DBLP:journals/tvcg/SrinivasanEBRE25} noted, attention does not have to be displayed; it could be used solely for analysis or evaluation.
This paper focuses on how the visualization might actively incorporate this data to help coordination, promote exploration, or support reflection; we refer to this process as \textit{revisualizing} the attention.

\paragraph{Presentation.}
The revisualization can be achieved through two fundamental presentation styles:
(a) \faLayerGroup~\textbf{embedding} the attention directly into the visualization; or
(b) \faColumns~\textbf{separating} the attention from the visualization in a view or overlay.
In collaborative contexts, the choice between approaches affects how easily participants can develop shared understanding of collective attention.
The presentation choice may also be left to each individual collaborator, and the actual implementation depends on whether the approach is data-aware or data-agnostic and whether it is displayed in 2D or 3D.

\paragraph{Situatedness.}
We also distinguish whether the attention is integrated into the \faGlobe~\textbf{world space} or the \faCamera~\textbf{camera space} of the display.
In \faGlobe~\textbf{world space}, attention visualizations remain fixed in the shared environment, preserving spatial context and enabling deictic reference (\textit{``look at that orange region''}).
In \faCamera~\textbf{camera space}, attention visualizations attach to each user's viewport, ensuring continuous visibility but sacrificing direct spatial correspondence.
This distinction matters particularly in immersive 3D where users occupy different positions with independent viewpoints.

\paragraph{Triggering.}
Following AAV~\cite{DBLP:journals/tvcg/SrinivasanEBRE25}, the triggering mechanisms for revisualization may significantly impact user experience:
\faSync\ \textbf{always-on}, which continuously displays attention (risking Midas touch or reinforcing loops);
\faHandPointer\ \textbf{explicit}, where users deliberately activate revisualization; and
\faBolt\ \textbf{implicit}, where the system automatically shows attention based on context.
In collaborative settings, different triggering mechanisms might apply to personal versus shared displays, and the ability to trigger might depend on user role.

\paragraph{Collaborative revisualization patterns.}

Several patterns emerge as particularly useful for collaborative settings:
\textit{Attention Heatmaps} show density of aggregate attention;
\textit{Coverage Maps} highlight areas that have received little attention; and
\textit{Focus Convergence} emphasizes where multiple users are focusing.
These address complementary collaborative needs identified in prior work on group awareness~\cite{DBLP:conf/cscw/GutwinG98} and collaborative visualization~\cite{Isenberg2011}: collective interest, thorough examination without redundancy~\cite{Sarvghad2016}, and shared discovery.
Alternative encodings are possible within this framework, including attention trajectories, gaze rays, attention glyphs, and volumetric fields; the choice depends on task requirements, team structure, and analytical goals.
For this investigation, we prioritize the three patterns as starting points that minimize visual complexity in immersive environments~\cite{Ens2021}.


\section{System: HeedVision}
\label{sec:impl}

We present \textsc{HeedVision}, a collaborative attention-aware visualization system for immersive analytics.
Within our design space, \textsc{HeedVision} implements attention that is \faLayerGroup~\textbf{embedded} in the \faGlobe~\textbf{world space} and triggered \faHandPointer~\textbf{explicitly}.
We additionally provide prototype implementations of the remaining three design space quadrants (Section~\ref{sec:quadrant-impl}, Figure~\ref{fig:implementations}).
We selected the world-space embedded configuration for \textsc{HeedVision} because embedded visualization preserves spatial context crucial for collaborative 3D exploration, world-space placement ensures both collaborators share a common reference frame for deictic communication, and explicit triggering gives users agency over when to consult attention information, reducing visual clutter during active exploration while supporting deliberate coordination.
These choices prioritize unobtrusive capture with user-controlled display, reflecting our goal of supporting (rather than disrupting) natural collaborative workflows.

\subsection{Attention Model}
\label{sec:attention-impl}

The core architecture features a distributed attention model that captures, records, and visualizes collective attention across multiple visualizations within a shared workspace.
Unlike single-user AAV systems, our model maintains separate attention maps for each collaborator on each visualization.
The attention model is implemented as a 3D voxel grid overlaid on each visualization, with individual voxels storing attention values for specific regions.
This supports both individual and aggregated attention maps, enabling users to review personal focus areas and identify regions of collective interest.

The system updates attention values in real time as collaborators interact.
The model follows principles of human short-term memory: attention accumulates when a user focuses on a specific region and decays over time as attention shifts elsewhere.
Only voxels that intersect with the actual geometry of a visualization register and accumulate attention data, creating a \textit{data-aware} approach that directly links attention to the visualization's geometric structure.
This same infrastructure can support \textit{data-agnostic} approaches through alternative capture mechanisms (see Section~\ref{sec:attention-capture}).

\begin{figure}[t]
    \centering
    \includegraphics[width=\linewidth]{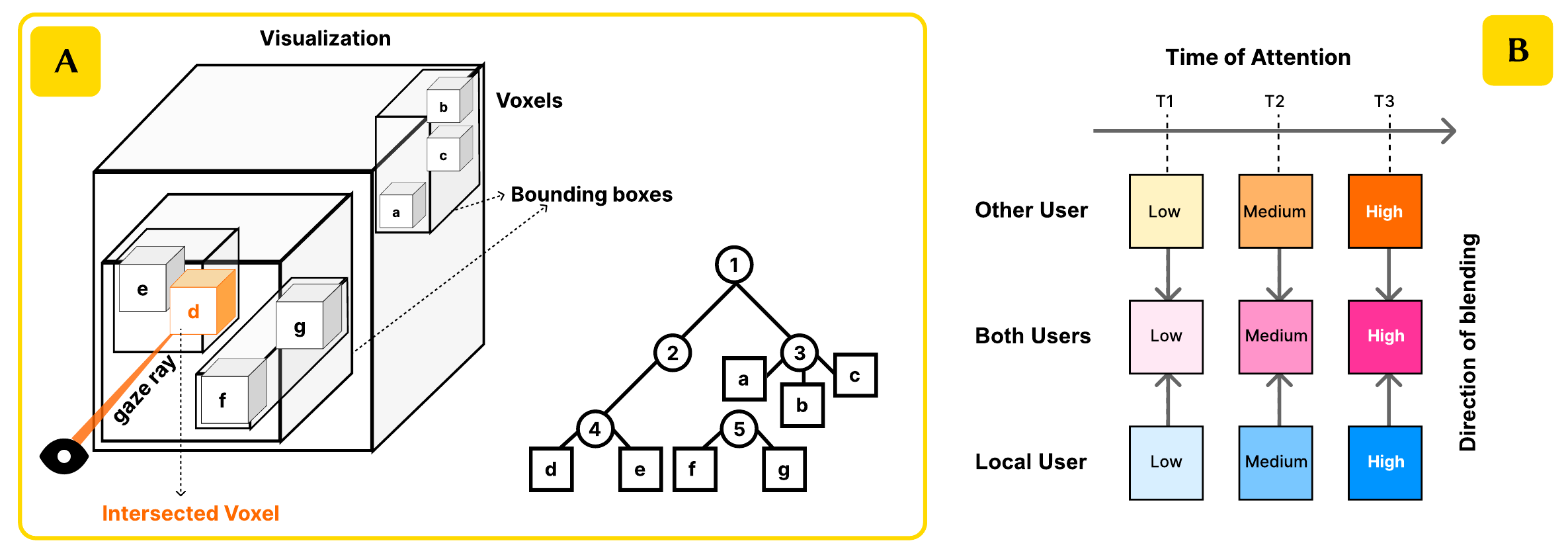}
    \caption{\textbf{Voxelization process.}
    (a) BVH~\cite{clark1976hierarchical} voxelization of a visualization, where the user's gaze ray intersects with the voxel grid, altering its state on intersection; (b) attention accumulation in individual voxels over time with color blending across two collaborators.}
    \label{fig:voxelization-process}
\end{figure}

\subsection{Voxelization}

At initialization, \textsc{HeedVision} converts the geometry of a visualization into a discrete grid of invisible voxels at configurable resolution.
These voxels serve as the fundamental units for capturing attention, becoming visible only when attention revisualization is triggered.
The voxel-based approach enables fine-grained spatial tracking: each voxel independently accumulates attention data, establishing a consistent spatial reference across users, simplifying detection of joint attention when multiple users focus on identical regions, and optimizing computation of attention coverage metrics.
We employ Bounding Volume Hierarchies (BVHs)~\cite{clark1976hierarchical} (Figure~\ref{fig:voxelization-process}) to accelerate voxelization by efficiently identifying mesh regions that intersect with voxel boundaries.

\subsection{Attention Capture}
\label{sec:attention-capture}

Given gaze information, we capture attention by casting a ray from the user's view position in their gaze direction, determining the nearest voxel intersection, creating a spherical \textit{region of influence} around it, and incrementing attention values for all voxels within that region with the center receiving the highest increment.
Using headsets without eye-tracking capabilities (such as the Meta Quest 3), we use head orientation as a proxy for user attention.
For a \textit{data-agnostic} approach, the algorithm would instead accumulate attention in every voxel the gaze ray passes through.

Head orientation introduces imprecision, as users often attend to peripheral regions.
Combined with our single-voxel region of influence (Section~\ref{sec:tech-impl}), this creates approximate attention maps.
However, this suffices for collaborative coordination where conveying \textit{general areas} of partner attention matters more than exact fixation points; this aligns with the strong head-gaze correlation documented in VR viewing studies~\cite{Sitzmann2018, David2018}.

\subsection{Triggering and Revisualization}
\label{sec:impl-triggering}

\textsc{HeedVision} implements an \faHandPointer~\textbf{explicit} triggering mechanism where participants press a controller button while pointing at a visualization to reveal collaborative attention patterns.
This prevents Midas touch~\cite{DBLP:conf/chi/Jacob90} while giving users control over when to consult the attention display.

For revisualization, \textsc{HeedVision} uses an \textit{opacity}-driven approach: each voxel adjusts its opacity based on accumulated attention, becoming more opaque as attention increases and remaining transparent when unattended or below a user-controlled threshold.
This creates a real-time coverage map that reveals where attention has been concentrated.
\textit{Color} distinguishes collaborators: each user is assigned a unique color, and when multiple users attend to the same voxel, their colors blend additively, effectively implementing a per-user aggregation that makes joint attention visually salient.
This color-based approach works well for small teams (2--4 members) but presents scalability challenges for larger groups, where alternative strategies like hierarchical grouping or role-based coloring may be needed (see Section~\ref{sec:limitations}).
A related consideration is potential conflict with data-driven color encodings: our current implementation assumes visualizations where color is not the primary data channel, but future work should explore alternative visual channels (texture, outline thickness) for color-heavy visualizations.
The attention visualization is directly embedded in the 3D world space, allowing users to explore attention patterns from any viewpoint while preserving spatial context.

\subsection{Implementation}
\label{sec:tech-impl}

\textsc{HeedVision} is built on open web technologies using React Three Fiber~\cite{React3FB} with \texttt{@react-three/xr} for WebXR support, running natively in modern browsers without specialized installation.
The collaborative infrastructure uses Spatialstrates~\cite{DBLP:conf/uist/BorowskiGBRKE25}, DashSpace~\cite{dashspace}, Webstrates~\cite{Klokmose2015webstrates}, and Varv~\cite{Borowski2022varv} for real-time cloud-based state synchronization, with an interactive Moveable marker system for positioning visualizations through drag and rotation.
The additional implementations of the remaining three quadrants (Section~\ref{sec:quadrant-impl}) are also built with React Three Fiber but operate in VR-only mode and use the BroadcastChannel API for local peer-to-peer synchronization between browser tabs without persistent cloud storage, enabling lightweight simulation of pair collaboration on a single device.

To maintain interactive frame rates, we made several trade-offs: attention capture and decay are computed only for the directly intersected voxel to avoid frame-blocking computations; the region of influence is restricted to a single voxel rather than surrounding neighbors; attention tracking operates at 10~Hz (below the 60~Hz rendering rate); and attention updates are batched and synchronized across devices at configurable intervals (200--500ms).
The primary bottleneck lies in network synchronization, as each update must be broadcast and merged across all clients.
Expanding beyond a single voxel to, say, a $3 \times 3 \times 3$ neighborhood would require $27\times$ more updates per frame, degrading responsiveness below 5~Hz on consumer hardware. Thus, a pragmatic tradeoff that sacrifices some spatial precision was necessary to maintain the responsiveness essential for fluid collaborative interaction.

\paragraph{Additional Design Space Implementations}
\label{sec:quadrant-impl}

\begin{figure}[t]
    \centering
    \includegraphics[alt={Four-panel figure showing implementations across all four design space quadrants. Top-left: World-space embedded (HeedVision) showing attention voxels directly overlaid on a terrain visualization. Top-right: World-space separated showing a miniature replica beside the main visualization. Bottom-left: Camera-space embedded showing 2D attention projections on the bounding box faces. Bottom-right: Camera-space separated showing a 2D attention panel in screen space. The top-left is the fully developed and evaluated system; the remaining three are proof-of-concept implementations.}, width=0.9\linewidth]{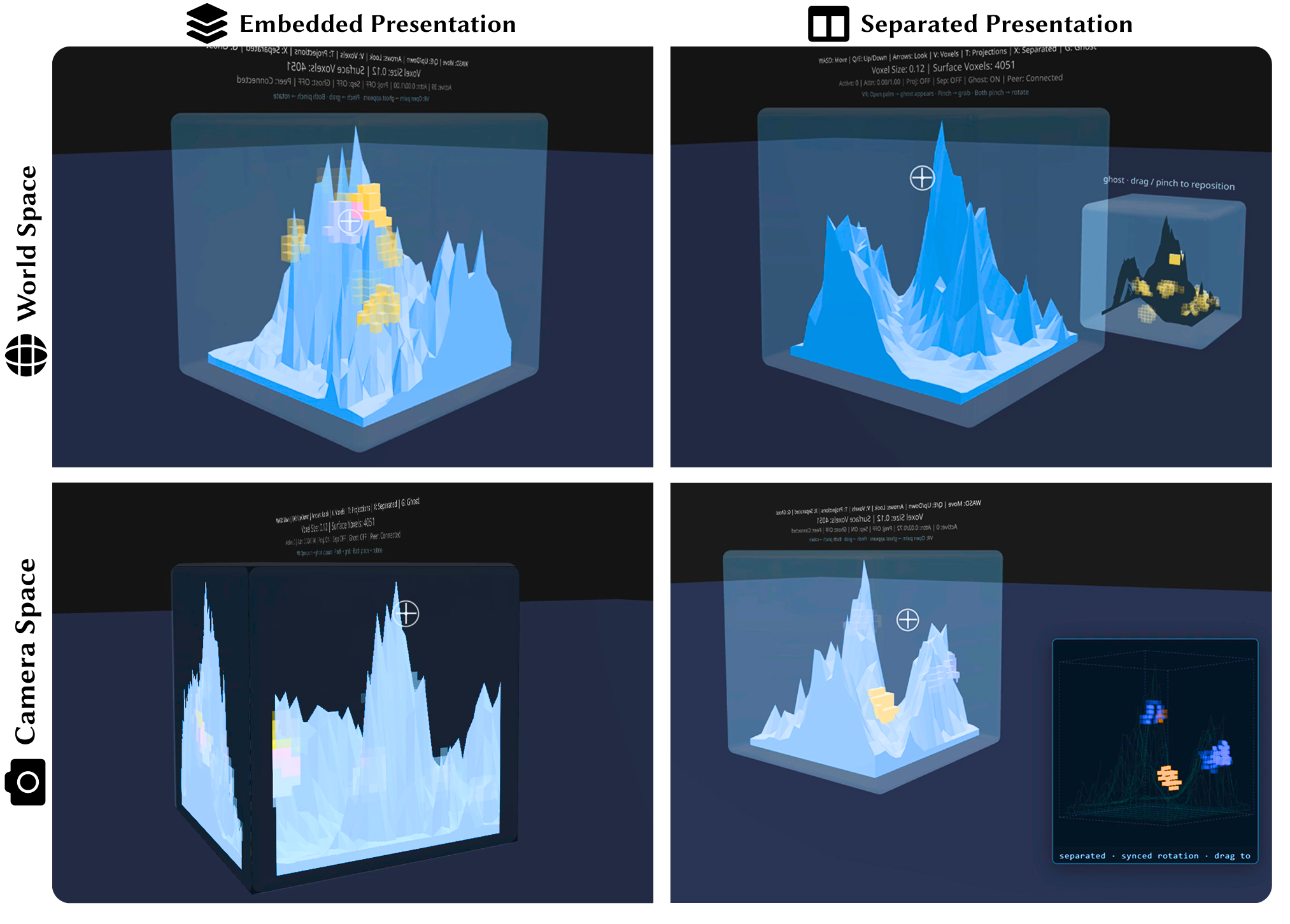}
    \caption{\textbf{Implementations across all four design space quadrants:}
    \textit{Top-left}: World-space embedded (HeedVision, fully evaluated).
    \textit{Top-right}: World-space separated.
    \textit{Bottom-left}: Camera-space embedded.
    \textit{Bottom-right}: Camera-space separated.
    The remaining three are prototype implementations sharing \textsc{HeedVision}'s attention model but operating in VR-only mode.}
    \label{fig:implementations}
\end{figure}

To demonstrate the breadth of our design space beyond \textsc{HeedVision}'s world-space embedded configuration, we provide implementations of the remaining three quadrants (Figure~\ref{fig:implementations}).
These implementations reuse the same attention model, voxelization pipeline, and capture mechanism described above, differing only in how attention is presented and situated.
Unlike \textsc{HeedVision}, which supports co-located AR with cloud-based synchronization, these implementations operate in VR-only mode and use the BroadcastChannel API to mimic pair collaboration locally; and have not undergone formal user evaluation.

\paragraph{World-space separated.}
A miniature replica of the visualization floats beside the main view and mirrors attention patterns in real time; users can reposition it via hand tracking or controller input to inspect collective attention from an independent vantage point.

\paragraph{Camera-space embedded.}
Attention is rendered as 2D projections overlaid onto the faces of the visualization's bounding box, computed from the user's current viewpoint, ensuring continuous visibility but sacrificing depth information.

\paragraph{Camera-space separated.}
A 2D canvas panel in screen space displays a top-down or rotated view of the attention map synchronized with the main visualization's orientation, providing an overview without occluding the 3D scene.

\section{User Study}
\label{sec:study}

To evaluate our proposed collaborative attention-aware visualization approach and the \textsc{HeedVision} implementation, we conducted a mixed-methods exploratory study.
Our goal was to understand how visualizing multiple users' attention patterns impacts coordination and coverage. 

\begin{table*}[htb]
\centering
\caption{\textbf{Participant demographics for collaborative pairs} (G1-G8) showing age, gender, vision, and their experience with immersive environments, co-located collaboration, and visual analytics ($N=16$).}
\label{tab:participants}
\resizebox{\textwidth}{!}{%
\begin{tabular}{c c c c c c c c | c c c c c c c}
\toprule
\multicolumn{8}{c|}{\textbf{Participant A}} & \multicolumn{7}{c}{\textbf{Participant B}} \\
\midrule
 & & & \textbf{Age Group} & \textbf{Vision} & \textbf{MR Exp.} & \textbf{Coloc. Exp.} & \textbf{VA Exp.} & & & \textbf{Age Group} & \textbf{Vision} & \textbf{MR Exp.} & \textbf{Coloc. Exp.} & \textbf{VA Exp.} \\
\midrule
\textbf{G1} & \faMars & P1 & 35-39 & \faGlasses & \faStar\faStar\faStar & \faStar\faStarHalf & \faStar & \faMars & P2 & 25-29 & \faEye & \faStar\faStarHalf & \faStar\faStarHalf & \faStar\faStar\faStar \\
\textbf{G2} & \faVenus & P3 & 25-29 & \faGlasses & \faBan & \faBan & \faStar & \faMars & P4 & 25-29 & \faGlasses & \faBan & \faStar & \faStar\faStarHalf \\
\textbf{G3} & \faVenus & P5 & 35-39 & \faEye & \faStar & \faStar\faStar & \faStar\faStar\faStar & \faMars & P6 & 25-29 & \faEye & \faStar & \faStar\faStar\faStar & \faStar\faStarHalf \\
\textbf{G4} & \faMars & P7 & 20-24 & \faGlasses & \faStar\faStar & \faStar\faStarHalf & \faStar\faStarHalf & \faMars & P8 & 25-29 & \faGlasses & \faStar & \faStar\faStar & \faStar \\
\textbf{G5} & \faVenus & P9 & 30-34 & \faEye & \faStar\faStar & \faStar\faStar\faStar & \faStar\faStar & \faMars & P10 & 30-34 & \faGlasses & \faStar\faStar\faStar & \faStar\faStar\faStar & \faStar\faStar\faStar \\
\textbf{G6} & \faMars & P11 & 25-29 & \faEye & \faStar\faStarHalf & \faStar\faStarHalf & \faStar\faStarHalf & \faMars & P12 & 20-24 & \faEye & \faStar & \faStar\faStar & \faStar\faStar \\
\textbf{G7} & \faVenus & P13 & 30-34 & \faEye & \faBan & \faStar\faStarHalf & \faStar\faStar & \faMars & P14 & 30-34 & \faGlasses & \faBan & \faStar\faStarHalf & \faStar \\
\textbf{G8} & \faMars & P15 & 25-29 & \faEye & \faStar & \faStar\faStar & \faStar\faStarHalf & \faVenus & P16 & 25-29 & \faGlasses & \faStar\faStar\faStar & \faStar\faStar\faStar & \faStar\faStar\faStar \\
\midrule
\multicolumn{8}{c}{\textit{Summary: $N$=16, 11\,\faMars\ 5\,\faVenus, ages 24--37 (M=28.9), 8\,\faGlasses\ 8\,\ \faEye}} & \multicolumn{7}{c}{} \\
\bottomrule
\vspace{0.5em}
\end{tabular}%
}
\vspace{0.5em}
\begin{minipage}{\textwidth}
\footnotesize
\textbf{Legend:}
\textit{Gender:} \faMars\ Male, \faVenus\ Female;
\textit{Vision:} \faEye\ Normal vision, \faGlasses\ Corrected-to-normal;
\textit{Experience:} \faStar\faStar\faStar\ Very experienced, \faStar\faStar\ Experienced, \faStar\faStarHalf\ Some experience, \faStar\ Novice, \faBan\ No experience;
\textit{Abbreviations:} \textbf{MR} = Mixed Reality, \textbf{Coloc}. = Co-located Collaboration, \textbf{VA} = Visual Analytics.
\end{minipage}
\end{table*}

\subsection{Participants}

We recruited 16 participants (11 male, 5 female) who formed 8 collaborative pairs (see Table~\ref{tab:participants}).
Participants were drawn from graduate-level Computer Science, Data Visualization, and HCI courses, ensuring sufficient background knowledge in data visualization.
Ages ranged from 24 to 37 (mean 28.9), and all had normal or corrected-to-normal vision with no self-reported color vision deficiency.
The participants varied in their experience with immersive environments, co-located collaboration, and visual analytics (from no experience to experienced).
None of the participants reported prior familiarity with the Mt.\ Bruno terrain dataset used in the study.

\subsection{Apparatus}

All participants used our implementation running on Meta Quest 3 headsets in a shared physical and virtual environment.
The system tracked head position and orientation to determine attention for each participant.
Pairs were co-located in the same laboratory space (6$\times$3$\times$3m), allowing for verbal communication.
Each participant's headset ran the \textsc{HeedVision} software on the Quest Browser.
The world-embedded attention display could be explicitly triggered using a controller button.

\subsection{Experimental Conditions}

We employed a 2$\times$2 within-participants design with two factors:

\begin{itemize}
\item\textbf{Visualization Type.} The 3D chart type used.
\begin{itemize}
        \item \textbf{Terrain:}
        A continuous 3D terrain visualization of Mt.\ Bruno's elevation with targets positioned along the surfaces of the entire volume.
        \item \textbf{Scatter:}
        A discrete 3D scatterplot with 100 uniformly distributed data points.
        We chose this distribution rather than realistic clustered data (through dimensionality reduction techniques like t-SNE or UMAP) to isolate CAAV's effect from coordination affordances that natural clusters might provide, thus creating a baseline where spatial structure alone offers no guidance for dividing labor.
    \end{itemize}

\item \textbf{Attention Visualization.} Attention display method.
    \begin{itemize}
        \item \textbf{No-CAAV (control):}
        No attention display.
        \item \textbf{CAAV:}
        World-embedded attention display.
    \end{itemize}
\end{itemize}

This produced four experimental conditions: \textit{Scatter $\times$ CAAV}, \textit{Scatter  $\times$ No-CAAV}, \textit{Terrain $\times$ CAAV}, and \textit{Terrain $\times$ No-CAAV}.
Conditions were counterbalanced across pairs using a Latin square design to mitigate learning effects.

\begin{figure}[htb]
    \centering
    \includegraphics[alt={Shows a pair of participants perform a treasure hunting task across two different visualizations, one with and the other without our CAAV.
    Left: Terrain Visualization with colored voxels over it. Right: Scatterplot visualization without voxels},width=0.7\linewidth]{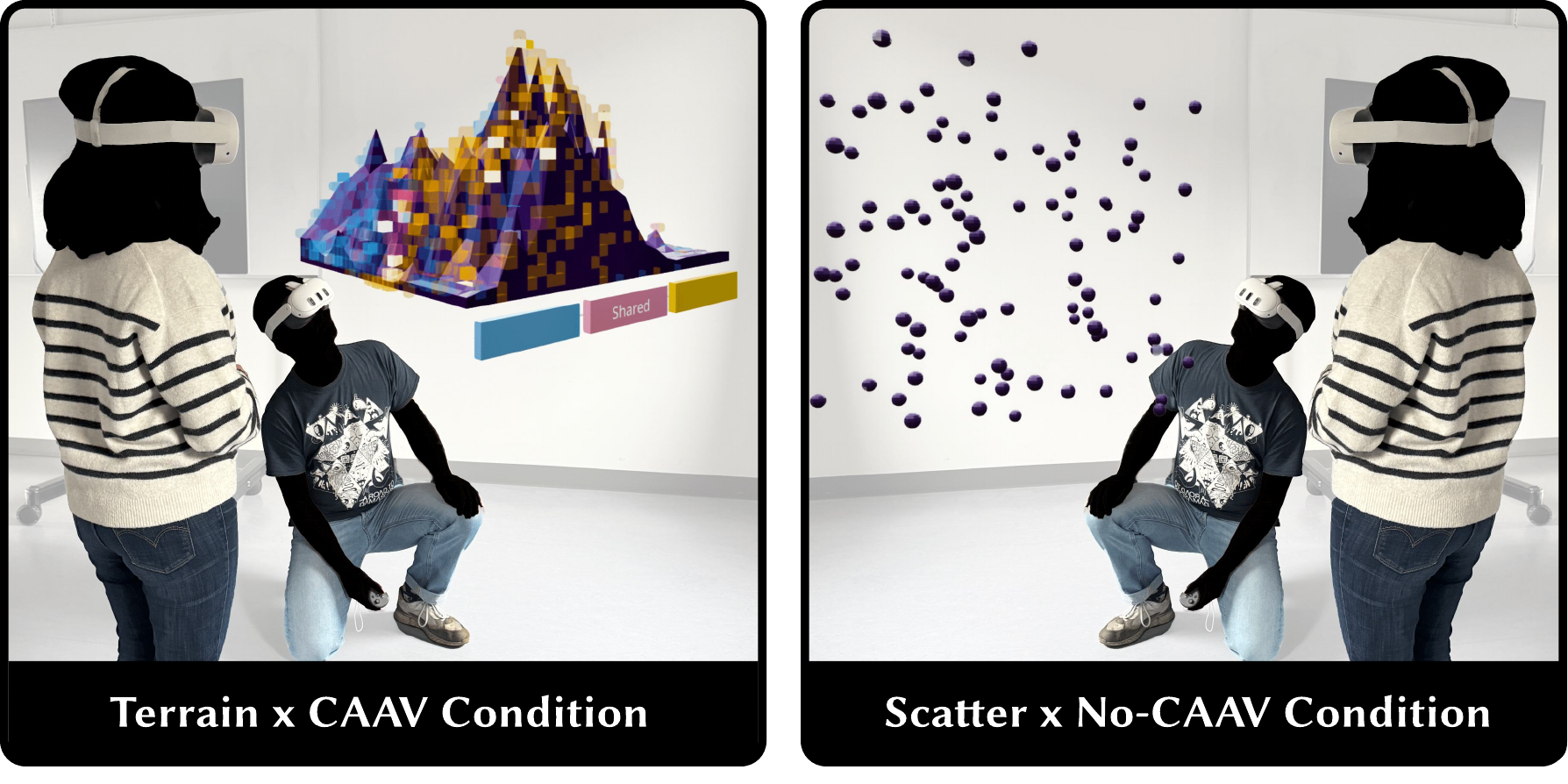}
    \caption{\textbf{Experiment with the treasure hunt task.}
    \normalfont
    A pair of participants perform the treasure hunt task.
    \textit{Left}: Terrain visualization with CAAV enabled (colored voxels visible). \textit{Right}: Scatterplot visualization without CAAV.
    All participants experienced both visualization types in both CAAV conditions.}
    \label{fig:experimental-setup}
\end{figure}

\begin{figure*}[t]
    \centering
    \includegraphics[alt={A multi-panel figure illustrating the HeedVision collaborative visualization system. Section A shows three sequential frames along a timeline: User A's initial view with blue voxels, User B joining and seeing both users' attention, and the collaborative state with blue (User A), yellow (User B), and pink (shared) voxels. Section B shows an enlarged detailed view with the gaze reticle, discovered targets, and color legend.}, width=\textwidth]{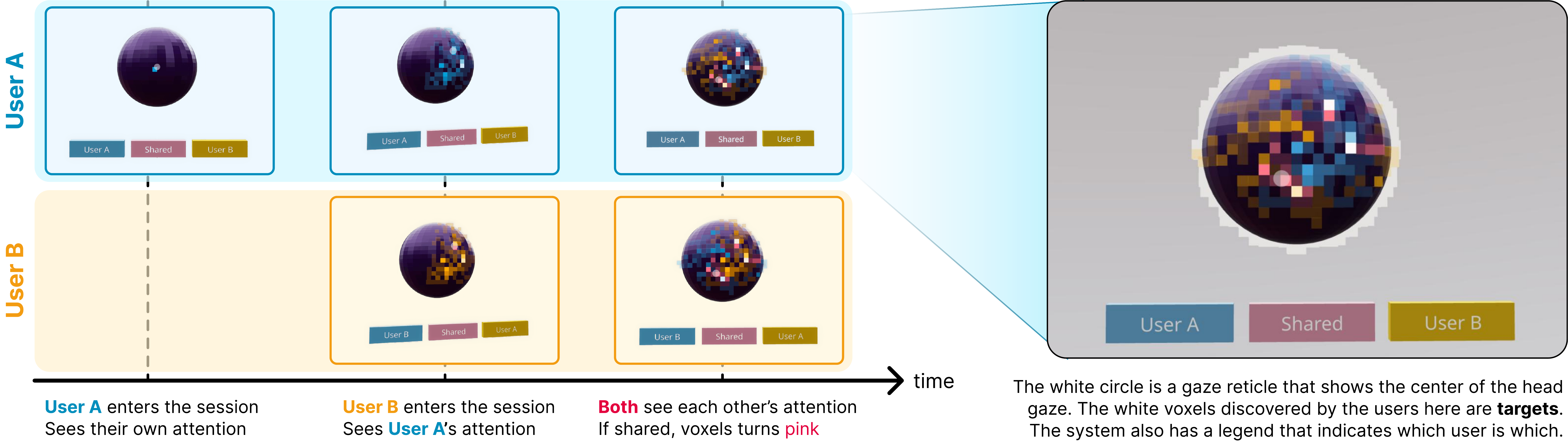}
    \caption{\textbf{HeedVision collaborative attention-aware visualization scenario}.
    (A) Timeline showing how attention visualization evolves as users collaborate:
    User A enters first and sees only their own attention; when User B joins, they see User A's accumulated attention; as both collaborate to discover targets, each sees their own and the other's attention with shared regions highlighted in pink.
    (B) Detailed view of the interface showing the gaze reticle (white circle indicating head gaze center), discovered targets (white voxels), and the legend indicating user identity through color coding.}
    \label{fig:hv-scenario}
\end{figure*}

\subsection{Task}

All pairs engaged in a collaborative visual search task: a treasure hunt (see Figures~\ref{fig:experimental-setup} and~\ref{fig:hv-scenario}).
Visual search tasks are well-established in immersive analytics research for evaluating spatial coordination and attention~\cite{10.1145/3773063, 10.5555/3141475.3141516}.
We designed this task as a simplified probe into key collaborative challenges (spatial navigation, shared attention, and division of labor), without domain- or task-specific complexity that might confound our analysis of attention-aware mechanisms.
During each 10-minute session, pairs searched for white target objects hidden within a 3D visualization.
These targets were initially invisible and only became visible (appearing to glow white) when a participant looked at them from the correct angle and within a 10~cm proximity.

In the CAAV condition, attention-based coloring supported collaboration: voxels changed color to indicate where participants had looked (blue for the local user, orange for the partner, and pink for areas explored by both).
Participants could toggle the visibility of these attention indicators using a controller button, enabling strategic exploration of unsearched regions and potentially enhancing collective search.

The number of targets was set to five percent of total voxels for each visualization type, with targets distributed pseudo-randomly across the visualization volume using a seeded random generator to ensure reproducibility.
For the terrain condition, targets were constrained to lie on or near the terrain surface; for the scatterplot condition, targets were distributed throughout the 3D volume surrounding the data points.
This distribution ensured that no single region was disproportionately target-rich, requiring pairs to explore the full visualization space.

\paragraph{Task Rationale.}

The ``treasure hunt'' was designed to isolate core coordination challenges while remaining simple enough to produce measurable outcomes within controlled conditions, aligning with established guidance for synthetic tasks~\cite{10.1145/1124772.1124774, wong2015evaluating}.
We chose manual spatial scanning because the coordination challenge arises from unknown target \textit{locations} distributed across a shared 3D space.
Unlike filter-based approaches that bypass spatial exploration entirely, manual scanning requires pairs to decide how to divide spatial coverage, making attention awareness directly relevant to task success.
This design parallels ``needle in a haystack'' scenarios common in exploratory data analysis~\cite{Grinstein2002InformationVisualizationVisual, Pirolli2005}.

Using both continuous (Terrain) and discrete (Scatterplot) visualizations allowed us to test CAAV across different data representation types.
The terrain provides natural landmarks and spatial continuity that may support coordination even without CAAV, while the uniform scatterplot lacks such affordances, allowing us to assess whether CAAV provides greater benefit in abstract environments.

\subsection{Procedure}

Participants provided informed consent, filled out a demographic form, and received a brief introduction to the system and task.
The order of CAAV and No-CAAV conditions was counterbalanced across pairs.
Before each condition, participants completed a brief training session to familiarize themselves with the XR environment, controls, and (for CAAV conditions) how to trigger and interpret attention visualizations.

Each pair completed the treasure hunt for all visualizations.
After each visualization, participants completed a brief questionnaire about their experience and coordination strategies.
Throughout the study, we recorded target counts, spatial coverage, attention overlap between participants, frequency of attention visualization triggering (CAAV condition), and verbal communication patterns.

After completing all conditions, participants were administered a System Usability Scale (SUS) questionnaire~\cite{brooke1996sus} and participated in a semi-structured interview about their experience.


\section{Results}
\label{sec:results}

We evaluated CAAVs to understand their impact on collaborative coordination during visual search tasks.
Our findings reveal that attention visualization's effectiveness depends on visualization context, with different patterns emerging across abstract discrete environments (scatterplots) and continuous spatial environments (terrain).
In scatterplot environments, pairs using CAAV showed improvements in coordination efficiency and reduced redundant exploration, while terrain environments revealed important boundary conditions for when attention visualization helps versus when existing environmental cues suffice.
These divergent results provide insights into the contextual factors that shape CAAV effectiveness.

\subsection{Task Performance}

We measured task performance through four complementary metrics.
\textit{Target Discovery} (percentage of targets found) captures task success directly.
\textit{Coverage} (percentage of space explored) reflects search thoroughness.
\textit{Coverage Efficiency} is the ratio of percent targets found to percent space explored; values above 1.0 suggest that pairs found targets at a rate exceeding what uniform random exploration would predict, though this interpretation assumes approximately uniform target distribution---a condition met by design for scatterplot but only approximate for terrain, where targets are constrained to the surface.
\textit{Gain Metrics} compare pair performance to individual baselines, revealing whether collaboration improved outcomes beyond individuals; target gain is computed as the pair's target count divided by the mean of the two individuals' target counts from the same condition, measuring the multiplicative benefit of working together. 

\begin{figure}[htb]
    \centering
    \includegraphics[alt={Grouped bar chart comparing target coverage and voxel coverage across four conditions. Scatter+CAAV: 78.5 percent targets, 76.5 percent voxels. Scatter+No-CAAV: 62.4 percent targets, 67.1 percent voxels. Terrain+CAAV: 26.0 percent targets, 24.6 percent voxels. Terrain+No-CAAV: 19.6 percent targets, 18.8 percent voxels.}, width=0.5\linewidth]{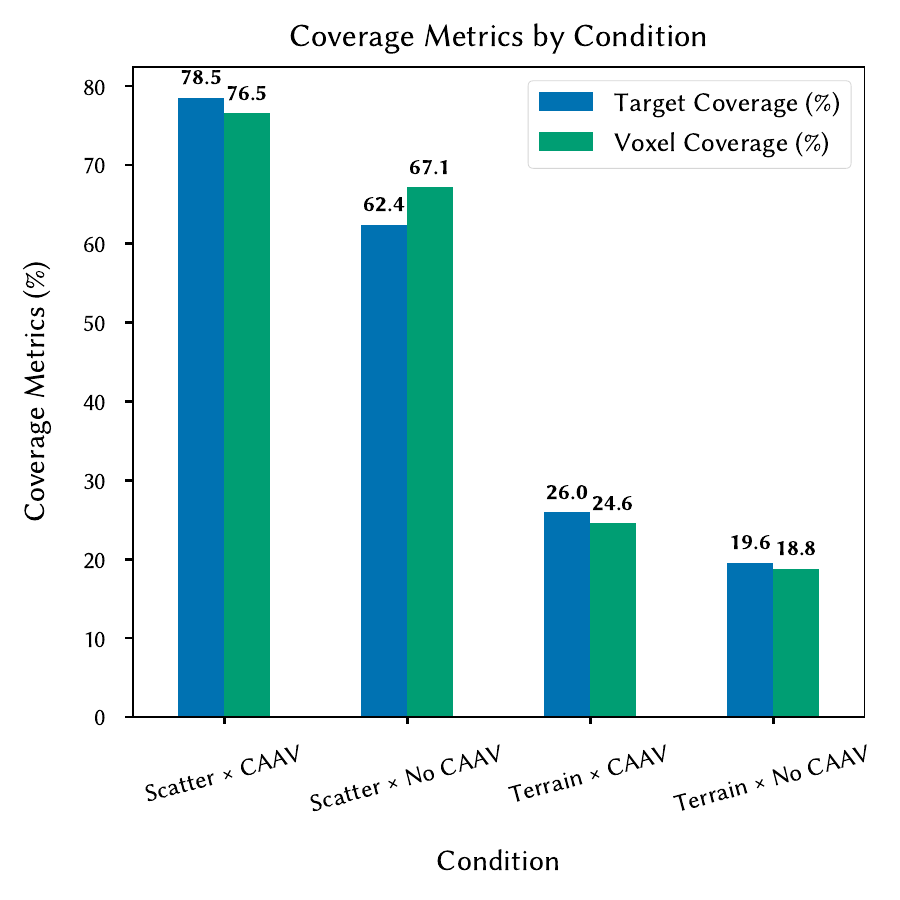}
    \caption{\textbf{Target and voxel coverage by condition.}
    Conditions are grouped by visualization type then by CAAV presence.
    CAAV conditions show higher target discovery rates, with scatterplot environments achieving substantially better coverage than terrain.}
    \label{fig:coverage_metrics}
\end{figure}

Pairs with CAAV showed improved performance across these measures.
For scatterplot, CAAV pairs found 78.5\% of targets compared to 62.4\% without attention visualization, while covering 76.5\% of the space versus 67.1\%.
In terrain, CAAV pairs found 26.0\% of targets versus 19.6\% without, and covered 24.6\% versus 18.8\% of the space.

Coverage efficiency exceeded 1.0 with CAAV in scatterplot conditions (1.027) but dropped to 0.920 without attention visualization, suggesting more effective search behavior with visual feedback.
Terrain conditions showed high efficiency regardless of CAAV presence (0.999 with CAAV, 1.027 without), suggesting that terrain geometry itself provides coordination cues.
The collaborative advantage was also evident in gain metrics: target gain was higher with CAAV in scatterplot conditions (1.33 vs.\ 1.19), while terrain conditions showed collaboration benefits regardless of CAAV presence (1.46 with CAAV vs.\ 1.19 without).
Within the scope of our two visualization types, these results suggest that attention visualization may provide greater benefit in abstract spaces lacking natural reference points.


\subsection{Spatial Coverage and Coordination}\label{sec:spatial-coverage}

To understand how pairs coordinated their exploration, we examined redundancy patterns alongside coverage.
\textit{Redundancy} measures duplicated exploration using normalized Shannon entropy: lower entropy implies more overlap, while higher entropy indicates diverse, non-redundant exploration.

Pairs using CAAV showed reduced redundancy, but patterns diverged by visualization type.
For scatterplot tasks, normalized redundancy dropped from 11.3\% (control) to 3.3\% (CAAV); this represents approximately a 70\% reduction.
For terrain tasks, however, CAAV \textit{increased} redundancy from 4.0\% to 14.9\%.
This reversal suggests that attention visualization interacts differently with environments that already contain natural coordination cues.

\begin{figure}[htb]
    \centering
    \includegraphics[alt={Bar chart showing normalized redundancy for four conditions. Scatter+CAAV: 0.033 (lowest). Scatter+No-CAAV: 0.113. Terrain+CAAV: 0.149 (highest). Terrain+No-CAAV: 0.040. CAAV reduced redundancy by 70 percent in scatterplots but increased it nearly fourfold in terrain.}, width=0.5\linewidth]{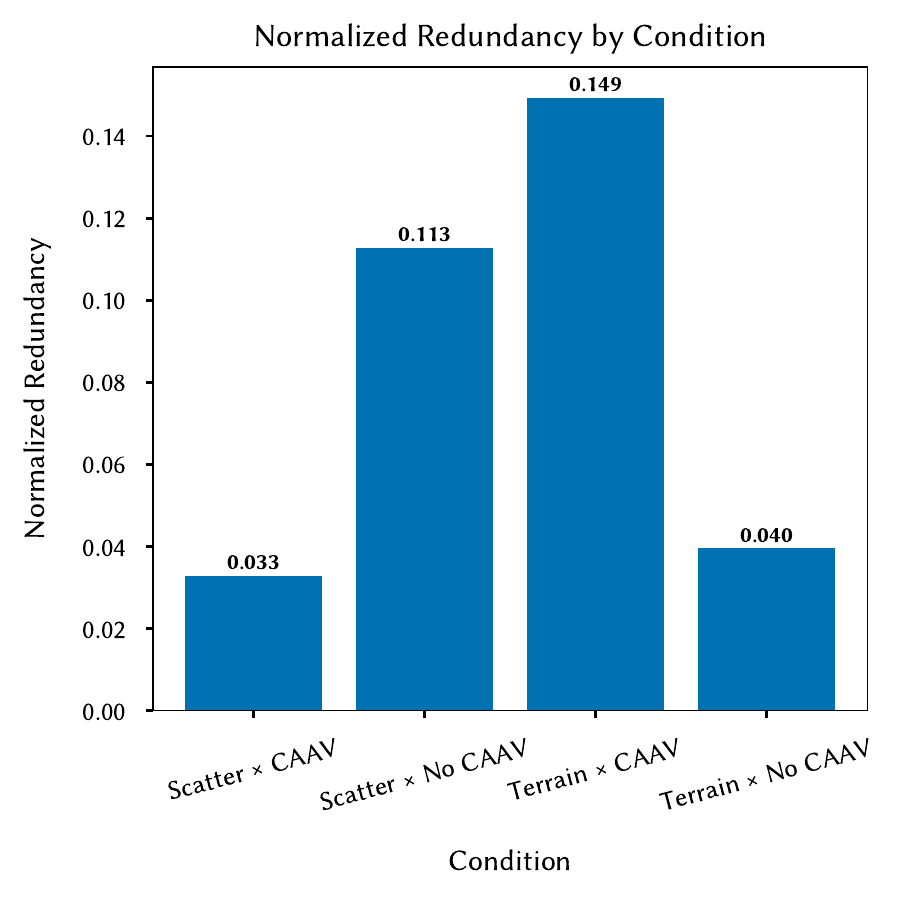}
    \caption{\textbf{Redundancy by condition.}
    Lower values indicate less overlap in exploration.
    CAAV substantially reduces redundancy in scatterplot environments but increases it in terrain, suggesting that attention visualization benefits depend on whether the environment already provides natural coordination cues.}
    \label{fig:redundancy}
\end{figure}

Behavioral observations help explain this pattern.
Participants in terrain + CAAV conditions often made deliberate second passes over explored areas, as if attention visualization made coverage gaps more salient and encouraged thoroughness over efficiency.
Terrain landmarks (peaks, valleys) naturally guided both participants toward the same prominent features; attention visualization then highlighted rather than prevented this convergence.

\begin{figure*}[htbp]
    \centering
    \includegraphics[alt={Three-panel terrain gradient-attention correlation analysis. Left: terrain gradient magnitude map. Center: aggregated attention heatmap. Right: overlay showing spatial correspondence. Pearson r=0.490, Spearman rho=0.561, both p<0.0001.}, width=0.9\textwidth]{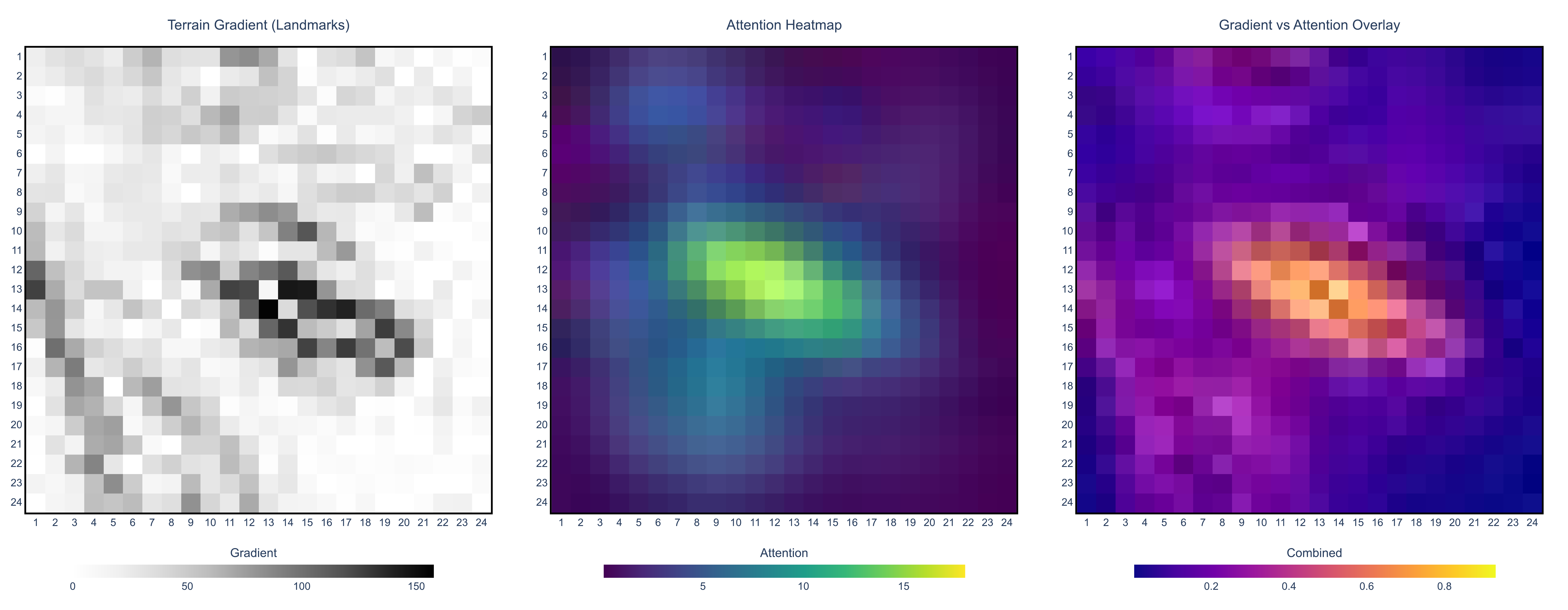}
    \caption{\textbf{Terrain gradient-attention correlation analysis.}
    \textit{Left}: Terrain gradient magnitude map showing topographic variation, where darker regions indicate steeper terrain features that serve as natural landmarks.
    \textit{Center}: Aggregated attention heatmap from all participants in terrain conditions, showing where collaborative attention concentrated.
    \textit{Right}: Overlay visualization showing the spatial correspondence between terrain gradients and attention patterns.
    Strong positive correlation (Pearson $r=0.490$, Spearman $\rho=0.561$, both $p<0.0001$) demonstrates that participants naturally converged attention on prominent landscape features (peaks, ridges, valleys), explaining the higher redundancy observed in terrain vs.\ scatterplot.} 
    \label{fig:terrain-gradient-attention}
\end{figure*}

Post-hoc correlation analysis provides empirical support for this interpretation.
We computed terrain gradient magnitude (the Euclidean norm of elevation differences between adjacent grid cells, where higher values indicate steeper slopes at prominent features like ridges, peaks, and valleys) across the terrain surface, then correlated these values with spatially-aggregated attention heatmaps from all terrain conditions.
Terrain gradients correlate strongly with collaborative attention (Fig.~\ref{fig:terrain-gradient-attention}): Pearson $r=0.490$ and Spearman $\rho=0.561$ (both $p<0.0001$).
Attention was also substantially more concentrated in terrain conditions (coefficient of variation: 2.231) than in scatter conditions (1.030). 

Together, these findings reveal the mechanism underlying terrain redundancy: prominent terrain features function as natural visual landmarks that draw collaborative attention regardless of CAAV presence.
This reframes our terrain findings not as CAAV failure, but as an indication that attention visualization's utility depends critically on the coordination affordances already present in the visualization environment.

Coordination efficiency reflected intentional strategies.
For scatterplot, participants described systematic division: \textit{``split into front/back faces for each of us''} [P7, G4] and \textit{``checking the color of CAAV, which matches our initial plan''} [P8, G4].
Without CAAV, pairs struggled and \textit{``had to double-check everything the other person looked at''} [P7, G4].

\subsection{Collaborative Strategies}

Through systematic thematic analysis~\cite{Braun2006} of post-study questionnaires, exit interviews, and observed behaviors, we identified six collaborative strategies (Table~\ref{tab:strategy_distribution}).
The analysis involved multiple coding passes: initial familiarization, systematic coding of coordination behaviors, pattern identification across conditions, and theme validation against quantitative behavioral data.
CAAV conditions enabled more sophisticated coordination patterns (\textit{spatial division}, i.e., dividing the environment into assigned regions; \textit{systematic search}, i.e., methodical, structured exploration; and \textit{visual feedback utilization}, i.e., leveraging voxel colors for awareness) while reducing reliance on \textit{verbal coordination}.
In abstract scatterplot environments, CAAV replaced verbal coordination entirely: P8 noted that \textit{``we have no verbal communication during the condition where we have [CAAV]''} [P8, G4].
Without CAAV, pairs developed compensatory strategies such as \textit{physical synchronization} (aligning markers in physical space).
Terrain showed the richest strategy diversity regardless of CAAV presence, with pairs combining \textit{landmark-based navigation} with other strategies.
This explained why CAAV provided less incremental benefit where the environment itself already supported effective coordination through its spatial structure.

\begin{table}[htb]
    \centering
    \caption{\textbf{Collaborative strategies by experimental condition}.
        Scatterplot (S) vs.\ Terrain (T); No CAAV (no) vs.\ CAAV (yes).
        Numbers indicate participant counts per strategy (by pair); participants may appear in multiple categories as pairs often employed multiple strategies simultaneously.
        CAAV conditions supported systematic search and visual feedback, reducing verbal coordination, while non-CAAV conditions relied more on verbal and compensatory strategies in abstract environments. For exact quotes by each pair in each category, please refer to the supplementary material.}
         \label{tab:strategy_distribution}
    \begin{tabular}{rcccc}
    \toprule
    \textbf{Strategy} & \textbf{S/No} & \textbf{S/Yes} & \textbf{T/No} & \textbf{T/Yes} \\
    \toprule
    \textit{Spatial Division} & 2 & 4  & 3 & 2 \\
    \textit{Landmark Navigation} & 2 & 0 & 3 & 5 \\
    \textit{Systematic Search} & 0 &4 & 1 & 2 \\
    \textit{Verbal Coordination} & 3 & 0 & 1  & 1 \\
    \textit{Visual Feedback} & 0 & 2 & 0 & 5\\
    \textit{Physical Synchronization} & 2 & 0 & 0 & 0 \\
    \bottomrule
    \end{tabular}
\end{table}

\subsection{Attention Visualization Usage}

To understand how participants utilized the CAAV interface, we logged toggle frequency and visibility duration.
Most participants kept the attention visualization active for approximately 75.2\% of session duration.
All participants used the toggle feature, indicating active management of the interface rather than passive acceptance.

\begin{figure}[ht!]
    \centering
    \includegraphics[alt={Timeline showing toggle behavior for 16 participants across 8 groups over 10-minute sessions. Colored bars indicate visualization ON; gaps indicate OFF. Most participants show predominantly colored bars with occasional short gaps, indicating they kept the visualization active for most sessions.}, width=0.8\linewidth]{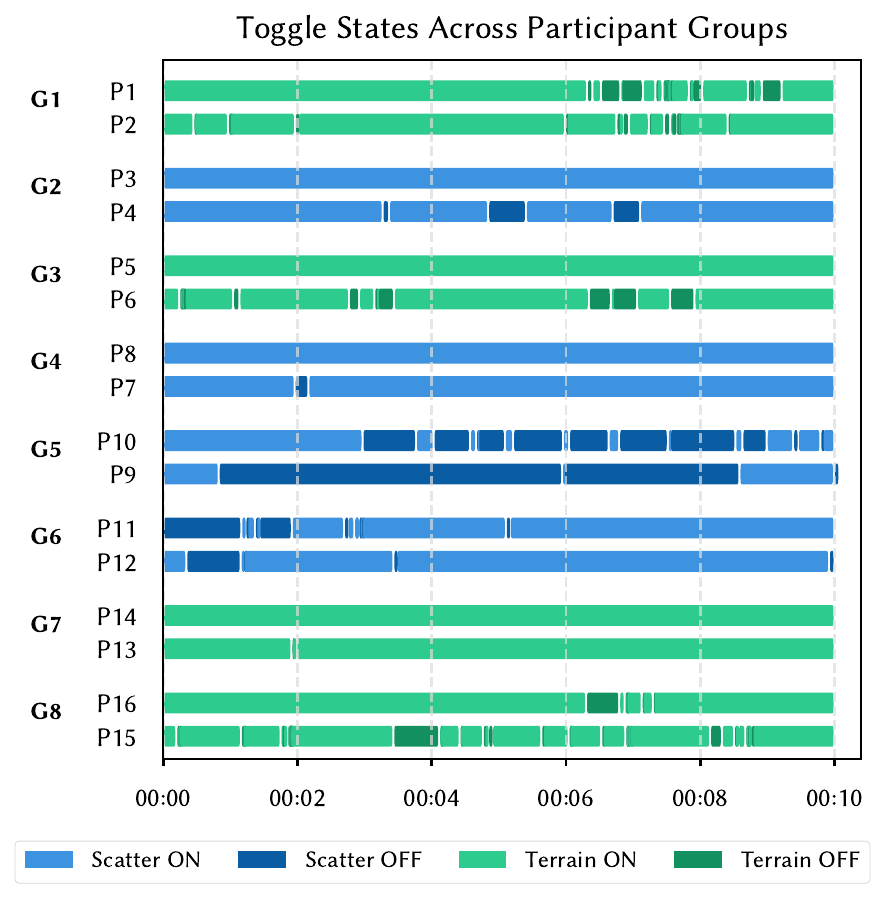}
    \caption{\textbf{Attention usage.}
    Toggle states across participant pairs showing when attention visualization was enabled (colored bars) versus disabled (gaps).
    Most participants kept the visualization active for the majority of their sessions, with strategic toggling rather than extended disuse.}
    \label{fig:toggle_states}
\end{figure}

Toggle behavior varied between individuals and conditions.
While some rarely disabled the visualization, others toggled it frequently.
P6 explained: \textit{``it was nice to be able to tell what the other person had seen... but having it turned off made the interface less overwhelming''} [P6, G3].
Even among frequent togglers, ``off'' periods were typically brief (averaging 4--5 seconds), suggesting participants valued the awareness cues despite visual complexity.

\subsection{Subjective Assessment}

Post-study ratings showed clear support for attention visualization: 93.8\% preferred CAAV conditions, with 81.3\% strongly agreeing that CAAV improved collaboration.
Preferences split between \textit{Scatter + CAAV} (50.0\%) and \textit{Terrain + CAAV} (43.8\%), with awareness of partner actions as the most cited reason.

The mean SUS score was 74.38 (SD = 19.79), above the 68 benchmark for good usability~\cite{brooke1996sus, Bangor_SUS}; 81.3\% of participants scored above this threshold, with only one outlier (P1, who preferred the non-CAAV condition).
NASA TLX ratings showed 24\% lower overall workload with CAAV (2.79 vs.\ 3.69 on a 1--7 scale), with reductions in mental demand (2.50 vs.\ 3.81), frustration (33\% lower), and temporal demand (24\% lower).
These improvements suggest that collaborative attention visualization provided clear subjective benefits for coordination, even in terrain conditions where coordination metrics showed mixed results.

\subsection{Qualitative Experiences}

Participants described CAAV as transforming collaboration from effortful coordination to intuitive teamwork: \textit{``seeing my partner's footsteps or breadcrumbs in real time.''}
The color encoding was frequently cited as helpful: \textit{``it was orange where they looked and got pink if I looked at it as well''} [P5, G3].
Communication dynamics shifted markedly: without CAAV, detailed spatial descriptions dominated; with CAAV, communication became strategic (``\textit{I think we're now at the phase where you complete my area and I complete your area}'' [P5, G3]), and several pairs reported no verbal communication at all.
Many participants treated the task as collaborative gameplay: \textit{``having the least pink as possible, and then having a nice blue side and a nice orange side with very little overlap''} [P6, G3].
Challenges centered on visual complexity over time, system latency (\textit{``The delay is too much''} [P2, G1]), and occlusion (\textit{``if a [face of a sphere] has blue [CAAV] and I have not explored it yet, it looks as if I have explored it''} [P12, G6]).
Despite these concerns, overall perception was positive, with participants envisioning a more implicit decay: \textit{``It would be nice if color of your attention would fade away [without noticing]''} [P8, G4].

\section{Discussion}
\label{sec:disc}

Our study was designed not to identify the optimal CAAV configuration, but to explore whether and how collaborative attention visualization affects coordination in immersive environments.
The findings reveal that CAAVs can improve spatial coordination, search efficiency, and task load distribution, though these benefits are strongly context-dependent.
Here we discuss their implications and limitations.

\subsection{Explaining the Results}

CAAVs address group awareness~\cite{DBLP:conf/cscw/GutwinG98} by making attention patterns visible, enabling what resembles ``stigmergy''~\cite{grasse1959}: coordination through environmental cues rather than direct communication.
The attention heatmaps function as ``read wear''~\cite{Hill1992} for collaborative visualization: a form of social navigation~\cite{Dourish1994} where users' traces become explicit and actionable.
This cognitive offloading (no longer needing to maintain a mental model of a partner's activities or verbally coordinate explorations) enabled participants to focus more on the analytical task itself, and was especially effective in scatterplot environments where pairs naturally distributed efforts with minimal verbal communication.

However, this mechanism worked differently across visualization types.
Our gradient-attention correlation (Section~\ref{sec:spatial-coverage}) suggests that prominent terrain features attracted both collaborators' attention regardless of CAAV presence, effectively providing built-in stigmergic cues; adding explicit attention visualization on top created redundant information channels, consistent with prior work on natural partitioning in continuous spatial environments~\cite{DBLP:conf/chi/ElmqvistTT08}.

\subsection{Generalizing the Results}

Based on our findings, CAAVs are likely beneficial when the visualization is \textbf{abstract and discrete} (lacking landmarks), collaborators must \textbf{divide a large search space}, the task requires \textbf{exhaustive coverage}, and \textbf{verbal descriptions of locations are difficult}.
Conversely, CAAVs may be less beneficial when the visualization has \textbf{inherent spatial structure}, or if the \textbf{visual complexity is already high}.
These conditions suggest CAAVs would be particularly valuable for high-dimensional data exploration (t-SNE, UMAP projections), molecular structure analysis, and 3D network visualization: domains where abstract spaces lack natural reference points, though further evaluation with domain-specific tasks is needed to confirm this.
Our random scatterplot represents an edge case, as real-world dimensionally-reduced data typically exhibits clustering; we chose random sampling specifically to create a ``pure'' abstract environment without emergent structure.
We hypothesize that CAAVs would remain beneficial for clustered data, though clusters could serve as partial reference that modulate the effect.

Our findings also extend beyond synchronous co-located settings: distributed collaboration could benefit even more from attention visualization, as it provides awareness otherwise completely absent.
Asynchronous collaboration could leverage attention decay to represent historical exploration, supporting handoffs between collaborators working at different times~\cite{Sarvghad2015}.

\subsection{Limitations and Future Work}\label{sec:limitations}

Several technical constraints bound our current implementation.
The voxel-based approach becomes expensive at high resolutions or with many users; adaptive resolution or GPU-accelerated tracking could address this.
Head orientation as an attention proxy offers less precision than eye tracking; we hypothesize our findings would transfer to eye-tracking implementations with potentially stronger effects.
Our color-per-user encoding limits scaling beyond four collaborators and may conflict with data-driven color encodings; alternative channels (hierarchical grouping, texture, animation, spatial distortion) could resolve both issues.
 
Our evaluation scope also constrains generalizability.
Although we provide prototype implementations for the remaining three design space quadrants, our formal evaluation focused solely on \textsc{HeedVision}'s world-space embedded configuration; the other configurations require further development and comparative evaluation.
Moreover, these proof-of-concepts use BroadcastChannel-based local synchronization rather than cloud-based infrastructure and operate in VR-only mode; maturing them for networked, co-located AR conditions remains future work.
The single task type, two-condition design, and pairs-only design do not capture the full complexity of collaborative analysis; future work should examine diverse tasks, larger teams, and selective sharing mechanisms that let collaborators choose when their attention is revealed.
Beyond spatial attention, future CAAVs could incorporate \textit{semantic} attention: tracking not just where users look but what data elements they attend to, providing richer collaborative awareness.


\section{Conclusion}

In this paper, we introduced Collaborative Attention-Aware Visualizations (CAAVs), extending attention awareness from individual to multi-user immersive analytics.
Our design space addresses three research questions: how to capture collective attention in real time (RQ1), how to record it over time (RQ2), and how to display it effectively (RQ3).
Through \textsc{HeedVision}, we demonstrated one viable configuration within this space (head-based capture, voxel-based recording with temporal decay, and explicit world-embedded display) and evaluated it with eight co-located pairs in augmented reality.

Our exploratory study provides initial evidence that CAAV effectiveness depends critically on visualization context.
In abstract, discrete environments such as scatterplots, CAAVs reduced redundant exploration by approximately 70\% and enabled emergent coordination strategies with minimal verbal communication.
In continuous terrain environments with natural landmarks, attention visualization sometimes interfered with existing coordination cues, suggesting that CAAV design must carefully consider the affordances of the visualization space itself.
These findings indicate that attention-aware collaboration may be most beneficial in abstract data spaces lacking natural reference points (such as high-dimensional projections, network visualizations, or dimensionally-reduced embeddings) while potentially offering less advantage in spatial data with inherent geographic structure.

Rather than prescribing a single optimal design, we emphasize the CAAV design space as a generative framework for future research.
Many configurations remain unevaluated: implicit triggering, camera-space presentation, data-agnostic recording, and integration with richer analytical tasks.
As immersive analytics continues to evolve toward multi-user environments, attention-aware approaches offer promising directions for enhancing collaborative sensemaking in spatial workspaces.


\section*{Supplemental Materials}
\label{sec:supplemental_materials}

All supplemental materials are available on OSF at \href{https://osf.io/frsp8/?view_only=0e5daa9f2e0246aaadb4601fbdf24ca1}{\texttt{osf.io/frsp8}} (anonymized), released under a CC BY 4.0 license.
In particular, they include the following:
(1)~Analysis files containing the final results of the quantitative task performance analysis, the NASA TLX and SUS analyses, and the consolidated qualitative analysis with participant quotes organized by theme;
(2)~Data files containing the collected experimental data, including spatial attention logs, toggle event streams, and participant responses from block-level and study-level surveys;
(3)~Study materials, including demographic forms, questionnaires, and additional figures for SUS scale; and
(4)~Source code for the HeedVision system and the proof-of-concept implementations of the remaining three design space quadrants.


\section*{Figure Credits}
\label{sec:figure_credits}

Unless specified below, all figures are credited to the authors.




\section*{Acknowledgements}

\paragraph{Use of Generative AI}

The three proof-of-concept implementations were developed with assistance from Claude Sonnet 4.5, which was used for debugging and documentation support.


\bibliographystyle{abbrv-doi-hyperref}
\bibliography{heedvision}



\end{document}